# ProfileSR-GAN: A GAN based Super-Resolution Method for Generating High-Resolution Load Profiles

Lidong Song, *Student Member, IEEE*, Yiyan Li, *Member, IEEE*, and Ning Lu, *Fellow, IEEE*

*Abstract*— This paper presents a novel two-stage load profile super-resolution (LPSR) framework, ProfileSR-GAN, to upsample the low-resolution load profiles (LRLPs) to high-resolution load profiles (HRLPs). The LPSR problem is formulated as a Maximum-a-Posteriori problem. In the first-stage, a GAN-based model is adopted to restore high-frequency components from the LRLPs. To reflect the load-weather dependency, aside from the LRLPs, the weather data is added as an input to the GAN-based model. In the second-stage, a polishing network guided by outline loss and switching loss is novelly introduced to remove the unrealistic power fluctuations in the generated HRLPs and improve the point-to-point matching accuracy. To evaluate the realisticness of the generated HRLPs, a new set of load shape evaluation metrics is developed. Simulation results show that: i) ProfileSR-GAN outperforms the state-of-the-art methods in all shape-based metrics and can achieve comparable performance with those methods in point-to-point matching accuracy, and ii) after applying ProfileSR-GAN to convert LRLPs to HRLPs, the performance of a downstream task, non-intrusive load monitoring, can be significantly improved. This demonstrates that ProfileSR-GAN is an effective new mechanism for restoring high-frequency components in downsampled time-series data sets and improves the performance of downstream tasks that require HR load profiles as inputs.

*Index Terms*—Generative Adversarial Networks, load profile generation, machine learning, non-intrusive load monitoring, super-resolution, synthetic data.

## I. INTRODUCTION

It is a common practice for utilities to downsample smart meter measurements from high-resolution (HR) to low-resolution (LR) (e.g., 15-, 30-, or 60-minute) [1]. This will significantly lower the costs for communicating, storing, and processing the data collected from millions of smart meters. Note that to be more specific, in this paper, HR refers to the sampling of data in 5-minute intervals or less.

In the past, smart meter data were mainly used for energy-related analysis, such as calculating monthly bills. A common practice is to record the total energy consumed in a sampling interval of 15 or 30 minutes, through which an average power consumption for the interval is computed. During this process, fast power variations within each sampling interval are lost, as shown in Fig. 1 (a) and (c). This is called the smoothing effect of signal averaging.

In recent years, two global activities, electrification in the transportation sector [2, 3] and decarbonization [4] in the energy sector, greatly expedited the integration of distributed energy resources (DERs), such as solar, wind, batteries, controllable loads, and electric vehicle chargers. However, uncertainties and variabilities inherent in renewable generation outputs and battery charging/discharging actions will lead to more frequent large and rapid power variations, causing circuit overloads and weakening voltage stability [5]. Consequently, for high-DER penetration distribution grids, where fast power variations are visible, it is increasingly important to use HR load profiles for conducting planning and operation studies such as quasi-static power flow analysis or non-intrusive load monitoring (NILM).

Recovering the HR data from the LR data is a data up-sampling process. In the past, interpolation is widely used to up-sample LR load profiles or patch the missing data. However, the main drawback of interpolation is that it cannot restore intra-interval power fluctuations.

In image processing, *super-resolution* (SR) [6] refers to the technique of generating HR images from LR images, as illustrated in Fig. 1 (c) and (d). A wide variety of deep-learning methods have been developed for image SR, for example, ultrasound imaging [7], line-fitting [8], and iris recognition [9]. Similarly, in audio signal processing, audio super-resolution (ASR) is used to recover the HR audio signals from the LR signals using deep-learning models [10].

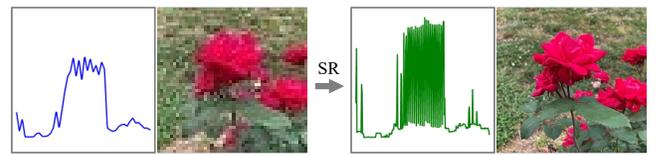

(a) A daily LR profile  (b) An LR image  (c) A daily HR profile  (d) An HR image

Fig. 1. An illustration of super-resolution problems. Generating *c* from *a* is an LPSR problem. Generating *d* from *b* is an image SR problem. Load profile data source: Pecan Street [11].

Motivated by image and audio SR, we define *load profile super-resolution* (LPSR) as a technique for generating realistic HR load profiles from LR load profiles. In power system

---

This research is supported by the U.S. Department of Energy's Office of Energy Efficiency and Renewable Energy (EERE) under the Solar Energy Technologies Office Award Number DE-EE0008770.

Lidong Song, Yiyan Li (*corresponding*), and Ning Lu are with the

Electrical & Computer Engineering Department, Future Renewable Energy Delivery and Management (FREEDM) Systems Center, North Carolina State University, Raleigh, NC 27606 USA. (e-mails: lsong4@ncsu.edu, yli257@ncsu.edu, nlu2@ncsu.edu).

applications, the development of LPSR is still in its infancy stage. In [12], Liu *et al.* developed super-resolution perception (SRP) for processing smart meter data. SRP combines Convolution Neural Network (CNN) with supervised learning based on Mean Square Error (MSE) loss. However, the MSE-based supervised learning algorithms can introduce unrealistic details and cause over-smoothing in the reconstructed HR data. This drawback has also been widely observed in image recovery studies [13].

Generative Adversarial Network (GAN) [14] based methods are widely used in solving image SR problems. Because the adversarial training between Generator and Discriminator can effectively capture the inherent probability distribution of the HR data, GAN-based methods can usually obtain more realistic SR results than the CNN or MSE based approaches [15].

In the power system domain, GAN has been used for generating load profiles [16, 17], wind farm outputs [18], simulating load forecasting uncertainty [19], synthesizing appliance power signatures [20], and appliance-level energy disaggregation [21]. However, to the best of our knowledge, using GAN-based methods for LPSR is still an uncharted area.

Thus, in this paper, we present a two-stage load profile super-resolution (LPSR) framework, ProfileSR-GAN. In the first stage, a GAN-based model is adopted to restore high-frequency components from the low-resolution load profiles (LRLPs). To reflect the load-weather dependency, aside from the LRLPs, the weather data is added as an input to the GAN-based model. The LPSR problem is formulated as a Maximum-a-Posteriori problem. To make the generated HRLPs more realistic, we use the method introduced in [13, 22] to construct the Generator loss function so that adversarial and feature-matching losses can be used to effectively recover the high-frequency components missed in the downsampling process. In the second stage, a *polishing* network is designed to connect to the GAN model to remove unrealistic power fluctuations from the generated HRLPs. A new set of load shape evaluation metrics is developed for evaluating the realisticness of the generated profiles and for comparing the performance with other state-of-the-art algorithms.

The main contributions of this paper are summarized as follows. Firstly, we formulated the LPSR problem as a Maximum-a-Posteriori problem that can be solved using GAN-based approaches. We added weather data as an input to the GAN model to reflect the load-weather dependency. Secondly, we proposed connecting a polishing network to the GAN model to remove unrealistic power fluctuations in the GAN generated HR load profiles. This significantly improves the ProfileSR-GAN performance on the point-to-point matching accuracy. Thirdly, we designed the performance evaluation metrics for evaluating the realisticness of the generated HR profiles. Our simulation results show ProfileSR-GAN outperforms the state-of-the-art algorithms. As an effective mechanism for restoring high-frequency components, ProfileSR-GAN can improve the performance of downstream tasks (e.g., NILM) that require HR load profiles as inputs.

The rest of the paper is organized as follows. Section II formulates the LPSR problem and introduces the ProfileSR-

GAN framework. Experimental results are presented in Section III. Section IV concludes the paper.

## II. Methodology

Let $P^{\text{HR}}$ represent the HR measurements with $N$ data points. $P^{\text{LR}}$ is the set of LR measurements down-sampled from $P^{\text{HR}}$ by averaging $\alpha$ continuous samples, where $\alpha$ is called the *scale-up factor*. Thus, $P^{\text{LR}}$ has $M$ data points and $M = N/\alpha$. The down-sampling process can be expressed as

$$P_m^{\text{LR}} = \frac{1}{\alpha} \sum_{n=\alpha(m-1)+1}^{\alpha m} P_n^{\text{HR}} + \eta_m$$
$$\forall m \in M, \forall n \in N \tag{1}$$

where $\eta$ represents noises caused by disturbances in the data acquisition process; $n$ and $m$ are the index of the HR and LR measurements, respectively.

As shown in Fig. 2, when a 1-minute load profile is down-sampled to 5-, 15-, and 30-minute load profiles, the high-frequency components will be filtered out because of the smoothing effect in signal averaging. The smoothing effect becomes more obvious when $\alpha$ increases. Compared with the 1-minute load profile, the 15- and 30-minute load profiles have lower load peaks and contain slower power variations. In addition, individual device on/off and cycling behaviors are no longer distinguishable. In the next few sections, we will introduce the ProfileSR-GAN for restoring the intra-interval power variations from LR load profiles.

### A. Load Profile Super Resolution Problem Formulation

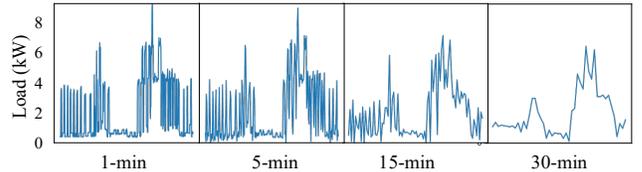

Fig. 2. Daily load profiles with different granularity. Assuming that the HR load profile has a sampling period of 1-minute. Then, $\alpha$ equals 5, 15, and 30 for the 5-, 15-, and 30-minute load profiles. Data source: Pecan Street[11].

Super-resolution algorithms originate from the image processing domain for processing 2-dimensional images. The 1-dimensional LPSR problem can be formulated as a Maximum a Posteriori (MAP) estimation problem as introduced in [23].

From the Bayesian rule, we have

$$p\left(P^{\text{HR}} \mid P^{\text{LR}}\right) = \frac{p\left(P^{\text{LR}} \mid P^{\text{HR}}\right) p\left(P^{\text{HR}}\right)}{p\left(P^{\text{LR}}\right)} \tag{2}$$

where $p(P^{\text{HR}}| P^{\text{LR}})$ is the conditional probability of $P^{\text{HR}}$ given $P^{\text{LR}}$, $p(P^{\text{LR}}| P^{\text{HR}})$ is the conditional probability of $P^{\text{LR}}$ given $P^{\text{HR}}$, and $p(P^{\text{HR}})$ and $p(P^{\text{LR}})$ are the prior probability of the HR and LR load profiles, respectively.

The objective function of an LPSR problem is to maximize the probability of the occurrence of $P^{\text{HR}}$ for a given $P^{\text{LR}}$. Using the MAP estimation method presented in [23], the estimated HR profile, $\hat{P}^{\text{HR}}$, can be obtained by

$$\hat{P}^{\text{HR}} = \arg\max_{\hat{P}^{\text{HR}}} \left(\log p\left(P^{\text{LR}} \mid \hat{P}^{\text{HR}}\right) + \log p\left(\hat{P}^{\text{HR}}\right)\right) \tag{3}$$

where $\hat{P}^{\text{HR}}$ is the estimated HR load profile.

In (3), the first term is commonly modeled using the Mean Squared Error between the actual LR profile, $P^{LR}$ and the LR

profile downsampled from the estimated HR one, $\hat{P}^{LR}$ [6, 24, 25]. The second term represents the prior knowledge of $P^{HR}$, which normally is modeled with a regulation factor to alleviate the ill-posedness of the SR problem [25]. Note that we refer to the prior knowledge as the information a learner already has before learning a new problem [26].

The traditional approaches only rely on minimizing the MSE loss[10, 12] to find a $\hat{P}^{HR}$. However, in practice, there is only one LR observation for a HR profile. As a result, solely relying on $\|\hat{P}^{LR}\text{-}P^{LR}\|_2^2$ can make the LPSR problem ill-posed, i.e., (3) may not have a unique solution [6]. For example, it is possible that the same LR profile can be obtained by down-sampling two different HR profiles.

Therefore, it is essential to constrain the solution space by introducing the prior knowledge of $P^{HR}$ into the SR problem formulation, as formulated by the second term in (3). In this paper, two types of prior knowledge are used. First, weather data and LR profiles are used as inputs to account for the known dependency of electricity consumption on the weather. Second, the two terms used in the generator loss function (refer to Section II.C): *adversarial loss* and *feature-matching loss*, represent prior knowledge of the shape characters extracted by the discriminator network from the actual waveforms.

### B. Generative Adversarial Network

A GAN model consists of two components: a generator network ($G$) and a discriminator network ($D$), as shown in Fig. 3. A latent vector $\mathbf{z}$, usually a Gaussian noise, is used as the input to generate the target output $G(\mathbf{z})$. Then, the generator output, $G(\mathbf{z})$, which is the generated data, and the real data, $\mathbf{x}$, are sent to $D$. The goal of $D$ is to distinguish which data sets are real and which are fake.

The training of a GAN model is an alternative and adversarial process: $G$ tries to generate samples $G(\mathbf{z})$ that can fool $D$; $D$ learns to identify $G(\mathbf{z})$ from $\mathbf{x}$ by assigning larger probabilities to $\mathbf{x}$ and smaller ones to $G(\mathbf{z})$. As introduced in [14], this process is formulated as a minimax game

$$\min_G \max_D V(D,G) = \mathbb{E}_{\mathbf{x} \sim p_{(\mathbf{x})}}[\log D(\mathbf{x})]$$
$$+ \mathbb{E}_{\mathbf{z} \sim p_{(\mathbf{z})}}[\log(1 - D(G(\mathbf{z})))] \quad (4)$$

where $V(D,G)$ is the reward function, $p_{(\mathbf{x})}$ and $p_{(\mathbf{z})}$ are the probability distributions of training data and latent vector, $\mathbb{E}$ is the expectation operator.

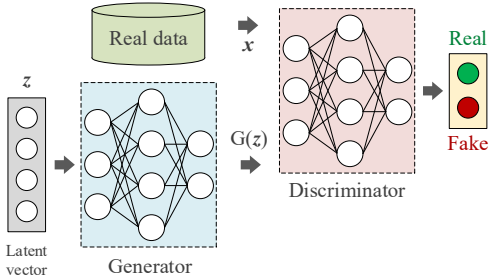

Fig. 3. The original generative adversarial network (GAN) model.

The GAN model learns the distribution-to-distribution mapping from the input to the output through the adversarial training of $G$ and $D$. In the LPSR problem formulation, we use the LRLPs obtained from smart meter measurements as inputs to learn the mapping from the implicit LRLP distribtution to the realistic HRLP distribution. By doing so, when a new LRLP is provided, the trained $G$ can generate a corresponding HRLP according to the learned LRLP-to-HRLP distribution mapping. In addition, we introduce additional loss terms and a polishing network to achieve a relatively low MSE, as shown in Section II.C.

### C. ProfileSR-GAN

In this section, we introduce architecture design, generator loss function selection, and polishing network loss function selection of the proposed ProfileSR-GAN framework.

#### 1) Network Architecture

As shown in Fig. 4, ProfileSR-GAN is a two-stage process. In the first stage, LR profiles and their corresponding weather data are used as inputs of the GAN-based model to generate HR profiles through adversarial training. In the second stage, a polishing network will remove unrealistic power fluctuations from the GAN generated HR profiles.

The generator network is a deep CNN. We firstly use convolution layers to extract high-level features from the input data. Then, we implement transpose convolution layers to recover the HR profiles. Inspired by [22], we use ReLU as the activation function. Residual blocks are inserted between two consecutive convolutional layers [27] to alleviate the gradient diminishing issue. We also adopt batch normalization following each convolutional layer [28] to enhance the training process.

The architecture summarized by Radford et al. [29] is used for constructing the discriminator network. The activation function is LeakyReLU. The discriminator network is trained to solve the maximization problem defined by (4). It contains four convolutional layers with an increasing number of kernels from 4 to 32. This allows us to compress the input profiles to high-level feature maps. Finally, the resulting feature maps will go through a fully connected (FC) layer and a sigmoid function to obtain the probability for real/fake classification. The polishing network is also a deep CNN bearing similar network structures as those of the generator, except that the two up-sampling transpose convolution layers are removed, and the number of kernels is reduced.

#### 2) Loss function design for the generator network

Let $\theta_G$ be the parameter of the generator network. The generator loss, $L_G$, is minimized to find an optimal $\theta_G$ by

$$\min_{\theta_G} L_G\left(G_{\theta_G}\left(P^{LR}\right), P^{HR}\right) \quad (5)$$

$$L_G = L_{cont} + \lambda_1 L_{advs} + \lambda_2 L_{feat} \quad (6)$$

where $L_{cont}$ is the content loss; $L_{advs}$ is the adversarial loss; $L_{feat}$ is the feature-matching loss; $\lambda_1$ and $\lambda_2$ are the weight coefficients.

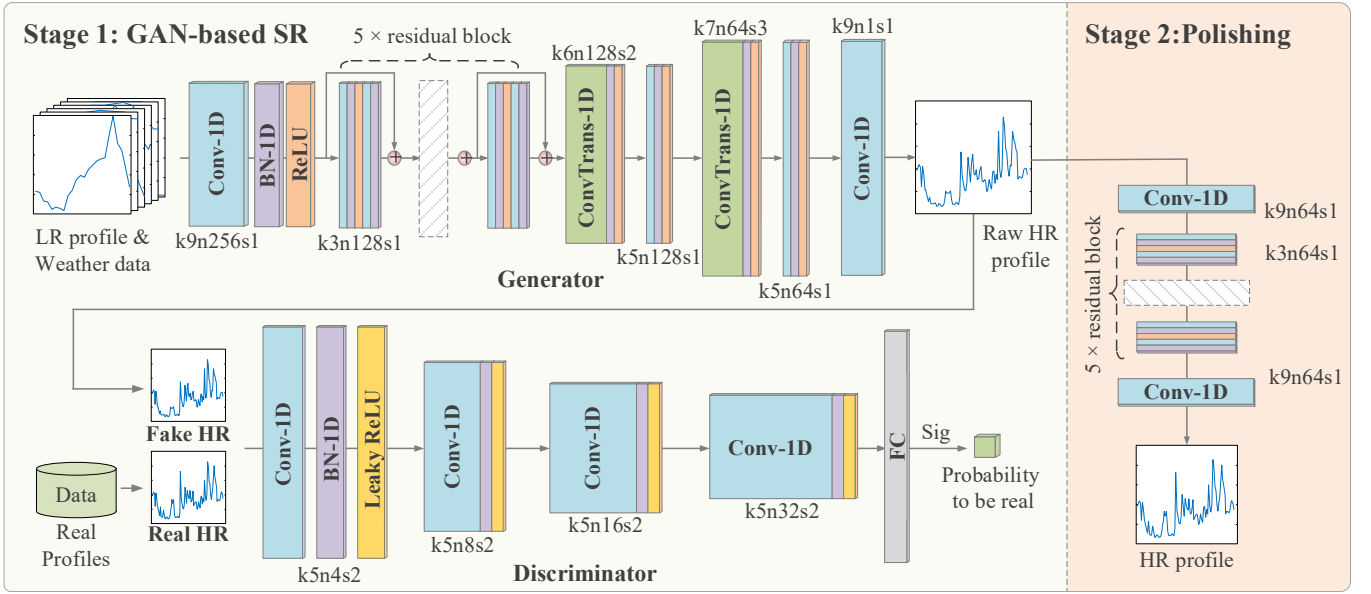

Fig. 4. The 2-stage ProfileSR-GAN architecture with corresponding kernel size (k), number of feature maps (n), and stride (s) indicated for each convolution layer.

To compute the content loss, $L_{cont}$, MSE is used to calculate the point-to-point distance between the generated and the ground truth HR profiles as

$$L_{cont} = \frac{1}{N} \left\| G_{\theta_G}(P^{LR}) - P^{HR} \right\|_2^2 \tag{7}$$

By minimizing the MSE, the generator is incentivized to find the maximum likelihood estimation of ground truth. However, relying solely on MSE-based $L_{cont}$ leads to over-conservative results in LPSR. As shown in Fig. 5, the generated HR profile is overly smooth, so it cannot restore high-frequency, large power variations. To resolve this issue, two loss terms, $L_{advs}$ and $L_{feat}$, are used in (6).

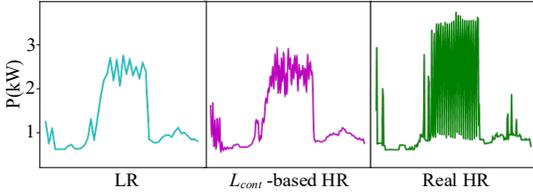

Fig. 5. The LR profile, the HR profile generated based solely on $L_{cont}$, and the real HR profile. All profiles are daily. Data source: Pecan Street [11].

The discriminator in GAN is trained to distinguish the fake from the real by minimizing the discriminator loss function, $L_D$, calculated as

$$L_D = -\left[ \log D_{\theta_D}(P^{HR}) + \log(1 - D_{\theta_D}(G_{\theta_G}(P^{LR}))) \right] \tag{8}$$

where $\theta_D$ is the parameters of the discriminator networks.

Let the second term related to $\theta_G$ in (8) be the adversarial loss $L_{advs}$. We have

$$L_{advs} = \log(1 - D_{\theta_D}(G_{\theta_G}(P^{LR}))) \tag{9}$$

By minimizing $L_{advs}$, the generator network favors solutions that cannot be distinguished as "fake" by the discriminator network to make the generated HR more realistic. Inspired by [14], we rewrite (9) as

$$L_{advs} = -\log(D_{\theta_D}(G_{\theta_G}(P^{LR}))) \tag{10}$$

to provide sufficient gradients and add robustness to the training process.

The feature-matching loss, $L_{feat}$, is defined as the distance between high-level feature maps extracted by the hidden layers of the discriminator network [30]. It is calculated as

$$L_{feat} = \sum_{j=1}^{J} \left\| \varphi_j \left( G_{\theta_G}(P^{LR}) \right) - \varphi_j \left( P^{HR} \right) \right\|^2 \tag{11}$$

where $\varphi_j(\cdot)$ represents the output of the $j^{th}$ intermediate convolution layer of the discriminator network, given real/fake HR profiles as inputs. $J$ is the number of intermediate layers involved in the loss function. As shown in Fig. 6, the extracted feature maps are quite different between real and fake profiles. Because high-frequency large power variations can be embedded in those hidden features, using $L_{feat}$ can train the generator to generate more realistic HR profiles.

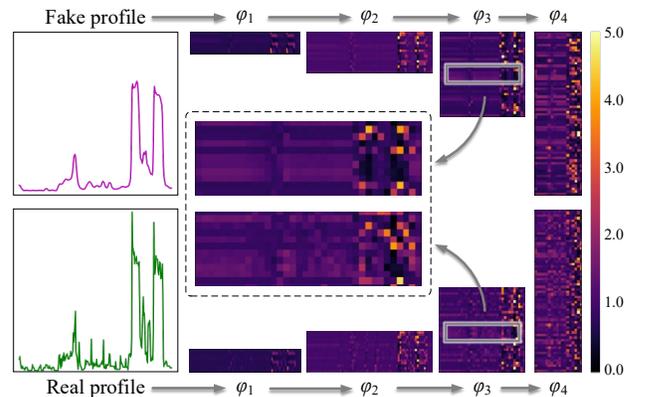

Fig. 6. Hidden feature maps extracted by the convolutional layers of the discriminator network. Load data source: Pecan Street [11].

Note that the output of the discriminator network is a binary classifier (yes/no) and cannot be directly used to calculate how close the generated load profile resembles the actual. However, one can compare feature values in each hidden convolutional layer when fake and actual profiles are examined. By minimizing $L_{feat}$, the generator will favor solutions that share similar features with the actual HR profiles to make the results more realistic.

### 3) Loss function design for the polishing network

Since the GAN model recovers the high-frequency components by reducing the adversarial loss and feature-matching loss, its performance in minimizing MSE loss is compromised. This is because point-to-point matching accuracy is sacrificed in exchange for the flexibility of generating more realistic details.

To resolve this issue, the loss function of the polishing network, $L_{pol}$, is designed to have two new loss terms: the *outline loss*, $L_{outl}$, and the *switching loss*, $L_{swit}$, so we have

$$L_{pol} = L_{outl} + L_{swit} \tag{12}$$

$$L_{outl} = \frac{1}{N} \left\| \xi_{\max}(\hat{P}^{\mathrm{HR}}) - \xi_{\max}(P^{\mathrm{HR}}) \right\|_2^2$$
$$+ \frac{1}{N} \left\| \xi_{\max}(-\hat{P}^{\mathrm{HR}}) - \xi_{\max}(-P^{\mathrm{HR}}) \right\|_2^2 \tag{13}$$

$$L_{swit} = \frac{1}{N} \left\| \xi_{\max} \left| \Delta \hat{P}^{\mathrm{HR}} \right| - \xi_{\max} \left| \Delta P^{\mathrm{HR}} \right| \right\|_2^2 \tag{14}$$

$$\Delta \hat{P}^{\mathrm{HR}} = \hat{P}^{\mathrm{HR}}(n+1) - \hat{P}^{\mathrm{HR}}(n), \quad \Delta P^{\mathrm{HR}} = P^{\mathrm{HR}}(n+1) - P^{\mathrm{HR}}(n)$$

where $\xi_{\max}$ is the max pooling operator [31] moving across the entire signal with a kernel size of $k_{max}$ at stride $s_{max}$ , $\Delta$ is the first-order difference operator. Note that $L_{outl}$ focuses on comparing the local peaks and valleys of the generated profile with the ground truth profile. This is also a proven effective solution used in solving image segmentation problems [32]. $L_{swit}$ focuses on comparing the change of load between two consecutive sampling intervals so that the load changing rates are similar to the ground truth profile.

Figure 7 shows an example of a daily load profile before and after polishing. Note that, the on/off of appliances normally lead to flat upper and lower boundaries instead of arbitrarily fluctuations. This is because an appliance usually runs at a relatively fixed power level. Together, $L_{outl}$ and $L_{swit}$ help flatten the unrealistic fluctuations to improve the point-to-point matching accuracy.

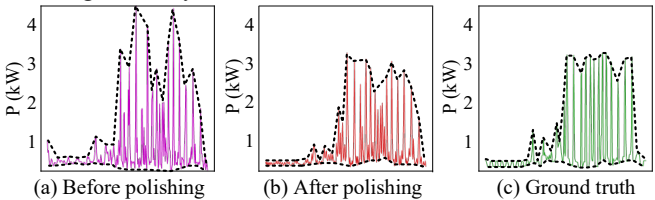

(a) Before polishing    (b) After polishing    (c) Ground truth

Fig. 7. An illustration of comparing the envelopes of the generated daily HR profiles (before and after polishing) with that of the actual daily load profile.

### D. Evaluation Metrics

In the image SR problem, evaluations of the restored HR images are usually based on a human-judgment-based measurement, called the Mean Opinion Score testing. However, in LPSR, substantial domain knowledge and expertise are required to visually distinguish whether a generated load profile is realistic. Therefore, we need a set of metrics for evaluating the quality of the generated HR load profiles.

MSE, as a point-wise comparison between two waveforms, is insufficient for comparing the shape of waveforms. For example, a minor time shift in waveforms can lead to a large

MSE, even though the two waveforms have the same shape. Therefore, in this paper, we introduce three shape-wise load profile evaluation metrics: Peak Load Error (PLE), Frequency Component Error (FCE), and Critical Point Error (CPE). The metrics are calculated as

$$PLE = \left| \max(G_{\theta_G}(P^{\mathrm{LR}})) - \max(P^{\mathrm{HR}}) \right| \tag{15}$$

$$FCE = \frac{1}{N} \left\| \mathcal{F}(G_{\theta_G}(P^{\mathrm{LR}})) - \mathcal{F}(P^{\mathrm{HR}}) \right\|_1 \tag{16}$$

$$CPE = \frac{1}{N} \left| \sum \mathcal{R}(G(P^{\mathrm{LR}})) - \sum \mathcal{R}(P^{\mathrm{HR}}) \right| \tag{17}$$

where $\mathcal{F}$ is the Discrete Fourier Transform operator and $\mathcal{R}$ is the Ramer Douglas Peucker operator [33].

PLE measures the peak load difference between the ground-truth HR profiles and the restored HR profiles by SR algorithms. PLE has a clear physical meaning and is important because accurately restoring the intra-interval peak load is critical for the distribution system operation and planning.

FCE measures the frequency-domain similarity between two profiles. Since the key challenge of LPSR is to restore the intra-interval, high-frequency components, FCE can compare the frequency-domain characteristics to evaluate the effectiveness of the SR algorithms.

Ramer Douglas Peucker algorithm simplifies a waveform in the time domain by eliminating non-critical points and keeping only shape-defining points [33]. Consequently, CPE measures the difference between the number of critical points of two waveforms to compare their similarity.

## III. SIMULATION RESULTS

To train and test the proposed model, we use the 1-min smart meter data set collected from 148 residential households in Austin, TX in 2015 by the PECAN Street association [11]. The hourly weather-related data set, including dry ball temperature, visibility, humidity, wind speed, sunrise and sunset time, are downloaded from [34] and then up-sampled to minute-level resolution by linear interpolation to pair with the LR load profiles.

The annual data are split into daily data. After excluding the days with missing data or abnormal data, we finally have 53,000 sets of daily load profiles. Those profiles are divided into two groups: 70% for training, 15% for validation, and 15% for testing. The 1-min data is down-sampled to 5-min and 30-min to obtain $P^{\mathrm{HR}}$ and $P^{\mathrm{LR}}$, so the scale factor $\alpha$ is 6. The Gaussian noise, $\eta$, has a zero mean with a variance of 0.01.

### A. Training Setup

Adam, an algorithm for first-order gradient-based optimization of stochastic objective functions introduced in [35], is used with momentum terms $\beta_1 = 0.99$ and $\beta_2 = 0.999$. The slope of the LeakyReLU is 0.2. Hyperparameters are tuned on the validation dataset listed in Table I. Deep neural network models are built in the PyTorch environment and trained on a single GPU of NVIDIA GeForce RTX 3080. The training time is approximately 10 hours.

To demonstrate the impact of introducing GAN-based components ($L_{advs}$ and $L_{feat}$), we design a CNN model that has the same network structure as that of the ProfileSR-GAN generator (see the upper left part in Fig. 4) as a controlled experiment. The CNN model is purely trained by the MSE loss defined in equation (7). The linear interpolation (LERP) method with a scale-up factor of $\alpha = 6$ is used as the benchmark case. Other SR approaches, including SRP [12], ASR [10] are also included in performance evaluation.

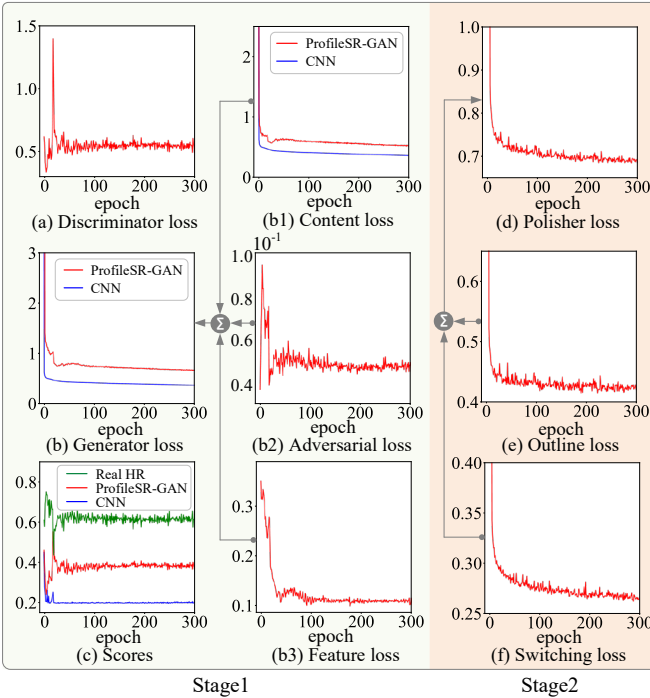

Fig. 8. Training curves. (a) discriminator loss of ProfileSR-GAN, (b) generator loss of ProfileSR-GAN and CNN, (b1) content loss of ProfileSR-GAN and CNN, (b2) adversarial loss of ProfileSR-GAN generator, (b3) feature loss of ProfileSR-GAN generator, (c) scores of real, fake HR profiles given by discriminator. (d) polishing loss, (e) outline loss, (f) switching loss.

TABLE I
HYPERPARAMETER SETUP

| Parameter | Value |
|---|---|
| Learning rate | 1e-4 |
| $L_{advs}$ weight - $\lambda_1$ | 0.05 |
| $L_{feat}$ weight - $\lambda_2$ | 0.5 |
| Batch size | 32 |
| Max pooling kernel size - $k_{max}$ | 3 |
| Max pooling stride - $s_{max}$ | 1 |
| Training epochs | 300(GAN) + 300(polishing) |

Figure 8 shows the loss plots in the two stages and scores given by the discriminator for the real and generated HR profiles during the GAN training process. The following observations are made for each training stage.

*1) Initialization (0 - 10 epoch)*

As shown in Fig. 8(a), initially, there is a sharp decrease of the discriminator loss. This means that the discriminator is learning the feature differences between the real and generated HR profiles quickly. Consequently, the discriminator starts to assign a higher score to the real profile and a lower score to the fake one, as shown in Fig. 8(c). Meanwhile, as shown in Fig. 8(b), although both CNN and ProfileSR-GAN have decreased generator losses, the generator loss of ProfileSR-GAN starts to bounce back quickly. This is because the decrease of content loss (see Fig. 8(b1)) is offset by the increase of the adversarial loss (see Fig. 8(b2)) and the feature-matching loss (see Fig. 8(b3)).

*2) Evolving (10-50 epoch):*

In this stage, for ProfileSR-GAN, the adversarial training allows the generator and discriminator to improve each other. The discriminator loss keeps decreasing, showing that the discriminator becomes more effective in identifying the real HR profiles from the fake ones. Note that despite the increasing generator loss, the performance of the generator continues to improve. This shows that guided by the discriminator, the generator is learning to achieve an optimal trade-off between *Achieving lower MSE error* and *generating more realistic profiles*. In other words, the content loss is sacrificed to lower the adversarial loss and feature losses. In contrast, the CNN model merely focuses on minimizing the MSE content loss.

*3) Balanced (after 100 epoch):*

After about 100 epochs, the generator and the discriminator of ProfileSR-GAN reach a balance in performance. There is no further decrease in discriminator loss, showing that the generator has the ability to generate realistic HR profiles to fool the discriminator. Meanwhile, the scores assigned to profiles by the discriminator are also stabilized: the score of a fake HR profile generated by the CNN model is around 0.2, much lower than those received by ProfileSR-GAN around 0.38). Note that the score of a real profile is approximately 0.62.

For the polishing network, both the outline loss and the switching loss drop sharply during the first 20 epochs and then stabilize at around $300^{th}$ epoch.

### B. LPSR results and Performance Evaluation

The LPSR results and metric values are summarized in Table II and Figs. 9 to 11. From the results, we have the following observations.

☐ Visual comparison of the generated daily load profiles. As shown in Figs. 9 and 10, the generated intra 30-min high-frequency components from ProfileSR-GAN are very similar to the ground truth profiles in terms of the magnitude of generated peaks, appliance cycling behaviors, and the envelopes of the load profiles. LERP and ASR fail to generate the intra 30-min power variations. In the daily profiles generated by SRP and CNN, the intra 30-min power variations are unrealistic with peak loads lower than that of the actual. This shows that ProfileSR-GAN can learn to exclude unrealistic components that are easy-to-be-identified-as-a-fake through the use of a discriminator.

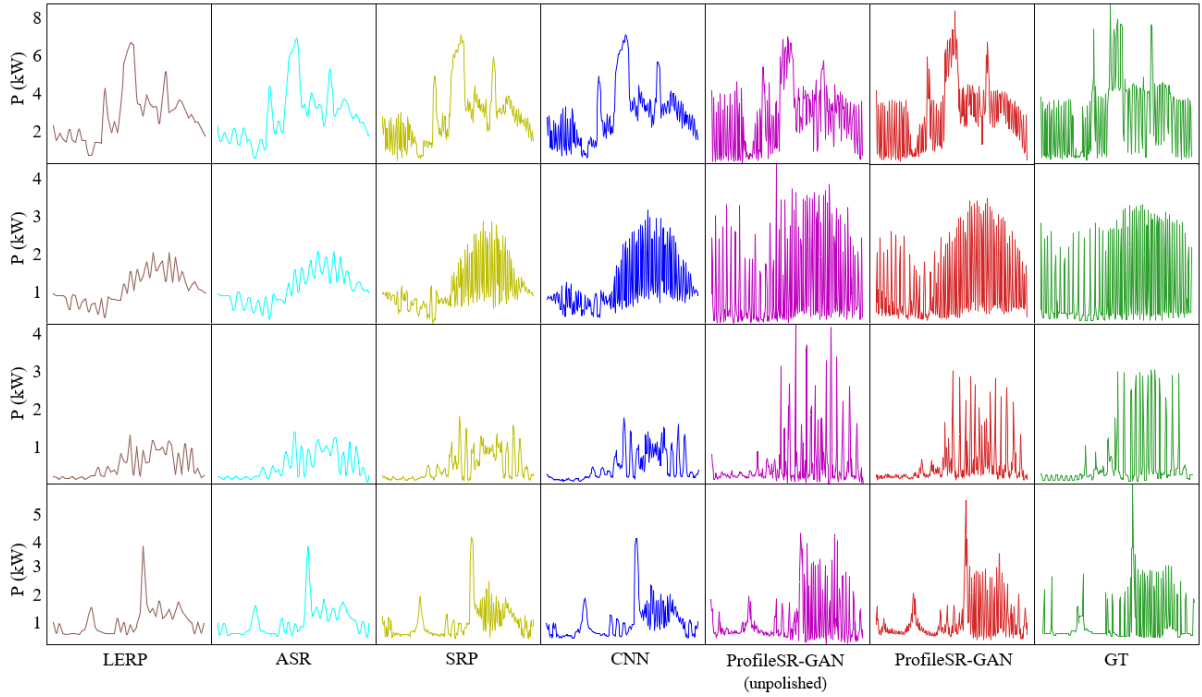

Fig. 9. LPSR results for generating daily load profiles (upsampling from 30-min to 5-min resolution). Lower resolution (LR), linear interpolation (LERP), convolution neural network (CNN), ProfileSR-GAN (unpolished), ProfileSR-GAN, and ground truth (GT). Data source: Pecan Street [11].

☐ Comparison of MSE: As shown in Table II, MSE-based methods outperform the GAN-based model in achieving the lowest MSE. This is reasonable since MSE is the only optimization objective those models need to take care of. However, only emphasizing point-wise MSE on averaging the point-to-point distance leads to smooth outputs. As shown in Figs. 9 and 10, such processes will filter out high-frequency components because recovering high-frequency detail is risking high point-to-point mismatch. A similar soothing effect has also been observed in the image SR problem and is reported in [13]. The results in Table II also shows that after adding a polishing network, there is an 18% improvement in MSE, showing the effectiveness of adding a polishing network to the ProfileSR-GAN architecture for improving point-wise accuracy

☐ Comparison of PLE: From Table II and Figs. 9 and 10, we can see that the magnitude of the peak in the ProfileSR-GAN generated profile is very similar to that in the ground truth curves. On the contrary, the peak load restored by LERP and MSE-based methods has a relatively large gap with the ground truth. Successfully restoring the load peaks is critical for distribution circuit analysis because load peaks often represent critical operating conditions.

☐ Comparison of FCE: From the spectrum plots in Fig. 10, we can see that the ProfileSR-GAN model can recover more high-frequency components than LERP and MSE-based methods.

☐ Comparison of CPE: As shown in Table II and Fig. 10, ProfileSR-GAN achieved the best performance in critical points matching. We can see that the simplified profile of ProfileSR-GAN is still very similar to the ground truth, indicating that they have similar profile complexity. The profiles generated by MSE-based methods and LERP, by contrast, have much fewer critical points.

☐ Comparison of the distribution of the shape-wise evaluation metrics on test set: As shown in Fig. 11, ProfileSR-GAN consistently outperforms LERP and other MSE-based models in terms of PLE, FCE, CPE while maintaining an acceptable MSE level.

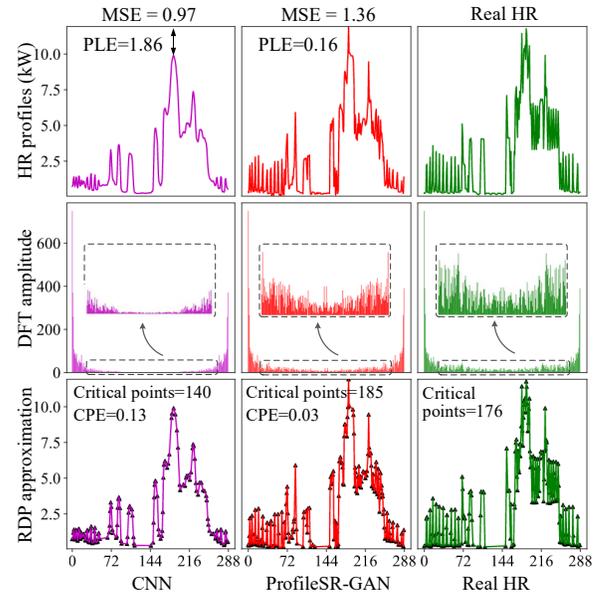

Fig. 10. Comparison of frequency components and critical point errors. (From top to bottom: HR profiles, the frequency components amplitude acquired by DFT, the simplified profile by RDP approximation. The critical points are shown with black markers. All load profiles are daily. From left to right: SR results of MSE-based CNN model, ProfileSR-GAN model, and real data). Data source: Pecan Street [11].

TABLE II
METRIC EVALUATION RESULTS

| SR method | | LERP | ASR | SRP | CNN | ProfileSR GAN- (unpolished) | ProfileSR GAN (polished) |
|---|---|---|---|---|---|---|---|
| MSE | mean | 0.55 | 0.44 | 0.42 | **0.41** | 0.61 | 0.51 |
| | Gain | / | 20% | 24% | 25% | -11% | 7% |
| PLE | mean | 1.38 | 0.99 | 0.92 | 0.91 | 0.86 | **0.73** |
| | Gain | / | 28% | 33% | 34% | 38% | 47% |
| FCE | mean | 7.22 | 5.83 | 5.36 | 5.38 | 4.81 | **4.65** |
| | Gain | / | 19% | 26% | 25% | 33% | 36% |
| CPE | mean | 0.65 | 0.41 | 0.29 | 0.31 | 0.26 | **0.25** |
| | Gain | / | 37% | 55% | 52% | 60% | 62% |

\* "Gain" represents the improvement w.r.t LERP baseline

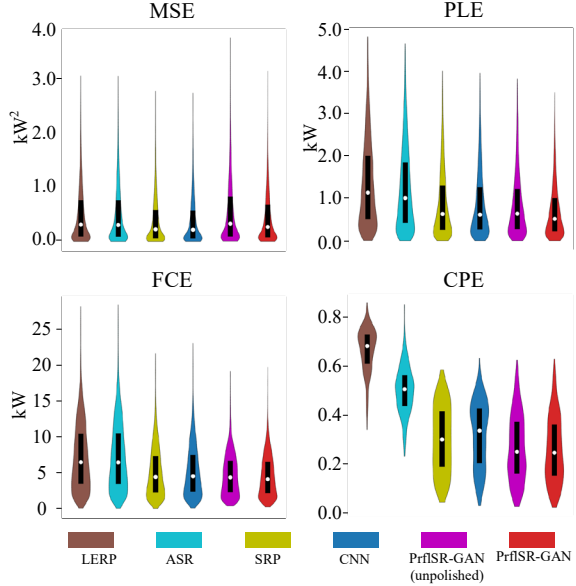

Fig. 11. Violin plots of the performance metrics. Medians are shown as the white markers.

## C. Distribution Visualization

The main advantage of using GAN-based methods is to generate HR profiles that have a similar probability distribution as the ground truth, which can hardly be achieved by MSE-based methods. To demonstrate this advantage, we select the generated daily HR profiles of a single house from April 1st to September 1st in 2015 to visualize the distribution, as shown in Fig. 12. We firstly use discrete Fourier transformation to extract the frequency-domain components of the generated HR profiles using different SR methods. Then, we use t-SNE [36] to reduce the dimension to 2-D for better visualization. We can see that the distribution of ProfileSR-GAN is the closest to the ground truth distribution. Meanwhile, Wasserstein distance is employed to quantify the distribution distances between the SR results and the ground truth. As shown in Table III, ProfileSR-GAN also achieves the best performance.

TABLE III
WASSERSTEIN DISTANCE OF DISTRIBUTIONS

| Metrics | SR models | LERP | ASR | SRP | CNN | ProfileSR -GAN |
|---|---|---|---|---|---|---|
| Wasserstein distance | X-axis | 4.560 | 4.518 | 3.391 | 3.506 | **0.679** |
| | Y-axis | 1.170 | 1.136 | 1.089 | 1.066 | **0.256** |

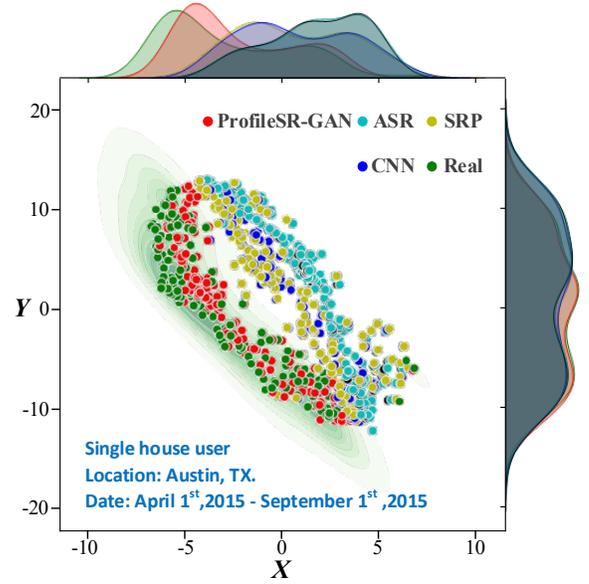

Fig. 12. 2D t-SNE visualization of the frequency components extracted from generated HR profiles using different SR methods. The data source for the actual load profile: Pecan Street [11].

## D. Performance Comparison under Different Scale-up Factor

To evaluate the impact of the scale-up factor $\alpha$, we compare the case of $\alpha$=6 (i.e. from LR-30min to HR-5min) with two other cases: $\alpha$=12 (i.e. from LR-60min to HR-5min) and $\alpha$=3 (i.e. from LR-15min to HR-5min). We keep the same ProfileSR-GAN network structure shown in Fig. 4 and alter only the stride of the transpose convolution layer of the generator network to cope with different $\alpha$ values.

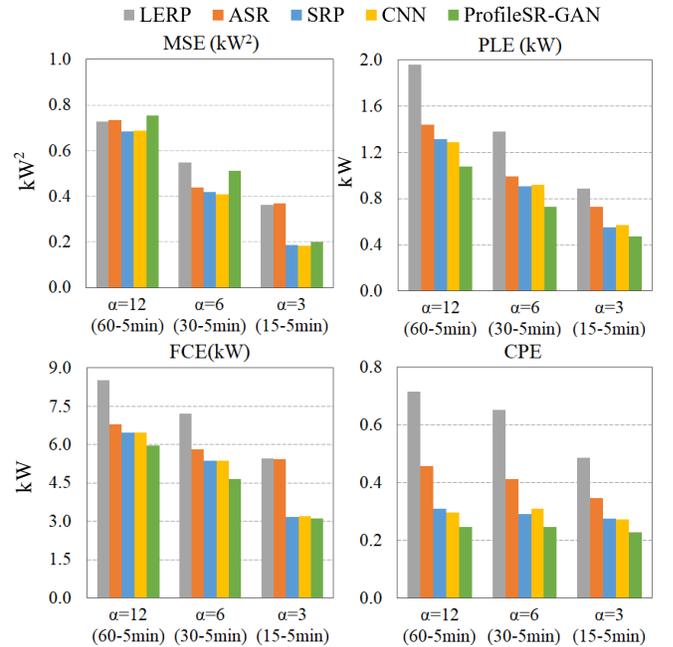

Fig. 13. Performance evaluation of cases with different scale-up factors.

As shown in Fig. 12, when $\alpha$ decreases, all SR methods perform better (i.e., four error metrics tend to reduce). This is expected because a smaller $\alpha$ represents a less ill-posed LPSR problem, making the problem easier to solve. In most cases, the MSE-based methods still outperform the GAN-based methods in MSE. We also observe that the MSE of ProfileSR-GAN significantly reduces when $\alpha$ decreases, showing that the LPSR problem becomes easier to solve when there are fewer points to recover in an interval.

Moreover, compared with LERP and MES-based learning, ProfileSR-GAN is more effective for larger $\alpha$ (e.g., $\alpha = 12$ and $\alpha = 6$). Because for a large $\alpha$, the LPSR problem is highly ill-posed: restoring the ground truth HR profile from the observed individual LR profile is challenging. However, the ProfileSR-GAN generator can approximate the distribution of the realistic HR profile dataset under the guidance of the feature-matching loss and adversarial loss. Therefore, it can generate HR samples that have a better chance to be realistic.

When $\alpha$ is small (e.g., $\alpha = 3$), the distribution approximation ability of the GAN-based method becomes less essential because each individual LP profile already contains enough information to recover the new missing points in an interval, making the MSE-based model better choices.

### E. Impact of Weather Data

As mentioned in Section III.A, weather data serves as part of the prior knowledge in this paper to enhance the GAN-based model performance. To assess the impact of the weather data on the performance of ProfileSR-GAN, we conduct a controlled experiment: we train two identical ProfileSR-GAN models, one with and the other without weather data. The performance metrics of the two models are summarized in Table IV. We can see that using the weather data as input achieves better performance in all four metrics.

TABLE IV
PERFORMANCE COMPARISON FOR QUANTIFYING THE IMPACT OF USING WEATHER DATA AS INPUT.

| Metrics | Weather Data | $\alpha$=12 (60 to 5min) | $\alpha$=6 (30 to 5min) | $\alpha$=3 (15 to 5min) |
|---|---|---|---|---|
| MSE | with | **0.782** | **0.512** | **0.201** |
| | without | 0.793 | 0.524 | 0.213 |
| PLE | with | **1.081** | **0.731** | **0.475** |
| | without | 1.145 | 0.765 | 0.491 |
| FCE | with | **5.964** | **4.652** | **3.125** |
| | without | 5.982 | 4.725 | 3.163 |
| CPE | with | **0.243** | **0.247** | **0.227** |
| | without | 0.251 | 0.252 | 0.259 |

### F. SR Results for Non-intrusive Load Monitoring

NILM methods are used to disaggregate electricity consumption from the smart meter level to the appliance level by capturing the unique signatures of each appliance. Because LPSR aims at restoring the high-frequency components of a down-sampled load profile, NILM is a natural downstream task and can be used to evaluate the effectiveness of LPSR. If the high-frequency load signatures can be recovered by LPSR, the NILM algorithm should be able to achieve better performance.

As shown in Fig. 14, the experiment includes four steps:

1) LPSR implementation. Five different LPSR methods are used, including the proposed ProfileSR-GAN and four other benchmarking methods, to upsample the 30-min aggregated LR load profiles back to 5-min HR load profiles.

2) NILM model training. The NILM models are trained using the real 5-min aggregated profiles. The outputs of the NILM are 5-min appliance level profiles. The trained NILM models can recognize appliances once provided with an aggregated load profile. In this paper, we will use three different NILM models: Denoising Autoencoder (DAE) [37], sequence to point model (Seq2Point) [38], and sequence to sequence model (Seq2Seq) [38]. These models are provided by the Non-intrusive Load Monitoring Toolkit (NILMTK) [39], which is an open-source NILM algorithm platform.

3) NILM model testing. After the NILM models are trained, they are fed with the HR profiles generated in step 1 to evaluate how well the appliance profiles can be recognized.

4) NILM result evaluation. The recognized appliance profiles in step 3 are compared with the ground truth to evaluate the NILM performance when using up-sampled load profiles by ProfileSR-GAN.

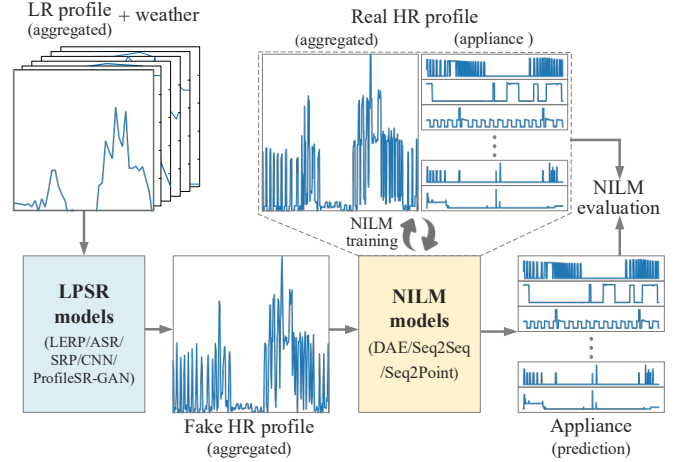

Fig. 14. Flowchart of the NILM experiments. Data source: Pecan Street [11].

Pecan Street data set paired with weather data from [34] is used to support the NILM experiments. We randomly selected four residential users from the data set. Each user has five appliances, i.e., air conditioner, electric furnace, fridge, dishwasher, and microwave. The data is one-month length from August 1st to September 1st, 2015 with 1-min granularity. We downsample the original 1-min aggregated profiles and appliance level profiles to 5-min as $P^{HR}$ and 30-min as $P^{LR}$. The first 20 days are used for training the NILM models, and the last 11 days for evaluation.

Two metrics are adopted to evaluate the NILM performance for each appliance. The first is the root mean square error (RMSE):

$$RMSE = \sqrt{\frac{1}{T}\sum_{t=1}^{T}\left(y_t - \hat{y}_t\right)^2} \qquad (18)$$

where $T$ is the profile length, $y_t$ is the actual power consumption of the target appliance at time $t$, and $\hat{y}_t$ is the corresponding NILM estimation. The second metric is the Overall Error (OE) [40], which measures the percentage power consumption mismatch between the NILM estimation and the ground truth.

$$OE = \left|\sum_t y_t \; / \; \sum_t Y_t - \sum_t \hat{y}_t \; / \; \sum_t \hat{Y}_t\right| \qquad (19)$$

Where $Y_t$ denote the actual aggregated power consumption of all appliances at time $t$, and $\hat{Y}_t$ is the corresponding NILM estimation. The calculation results are shown in Table V.

TABLE V
PERFORMANCE COMPARISON OF DIFFERENT NILM ALGORITHMS

| Appliance | Metrics | Root mean square error (kW) | | | | | Overall error $(10^{-1})$ | | | | |
|---|---|---|---|---|---|---|---|---|---|---|---|
| | Houses | LERP | ASR | SRP | CNN | ProfileSR-GAN | LERP | ASR | SRP | CNN | ProfileSR-GAN |
| Air-conditioner | 1 | 1.063 | 1.049 | 1.024 | 1.016 | 0.969 | 0.780 | 0.883 | 0.403 | 0.430 | 0.346 |
| | 2 | 1.216 | 1.201 | 1.071 | 1.103 | 1.048 | 2.417 | 2.303 | 1.587 | 1.917 | 0.758 |
| | 3 | 1.348 | 1.350 | 1.134 | 1.199 | 0.922 | 4.949 | 4.925 | 2.620 | 3.004 | 1.102 |
| | 4 | 0.793 | 0.809 | 0.710 | 0.727 | 0.679 | 0.713 | 0.790 | 0.404 | 0.453 | 0.104 |
| | **mean** | 1.105 | 1.102 | 0.985 | 1.011 | **0.905** | 2.215 | 2.225 | 1.253 | 1.451 | **0.578** |
| Fridge | 1 | 0.105 | 0.106 | 0.105 | 0.105 | 0.106 | 0.145 | 0.155 | 0.128 | 0.144 | 0.040 |
| | 2 | 0.071 | 0.074 | 0.074 | 0.078 | 0.067 | 0.325 | 0.285 | 0.233 | 0.261 | 0.160 |
| | 3 | 0.102 | 0.104 | 0.089 | 0.088 | 0.084 | 2.703 | 2.712 | 1.258 | 1.539 | 0.482 |
| | 4 | 0.078 | 0.079 | 0.078 | 0.079 | 0.078 | 0.278 | 0.335 | 0.138 | 0.153 | 0.236 |
| | **mean** | 0.089 | 0.091 | 0.087 | 0.087 | **0.084** | 0.863 | 0.872 | 0.439 | 0.524 | **0.230** |
| Electric furnace | 1 | 0.118 | 0.116 | 0.106 | 0.106 | 0.090 | 0.329 | 0.362 | 0.162 | 0.138 | 0.130 |
| | 2 | 0.063 | 0.064 | 0.057 | 0.058 | 0.056 | 0.687 | 0.634 | 0.501 | 0.644 | 0.474 |
| | 3 | 0.299 | 0.299 | 0.220 | 0.243 | 0.201 | 0.571 | 0.569 | 0.689 | 0.654 | 0.349 |
| | 4 | 0.071 | 0.073 | 0.066 | 0.067 | 0.064 | 0.055 | 0.057 | 0.049 | 0.064 | 0.021 |
| | **mean** | 0.138 | 0.138 | 0.112 | 0.119 | **0.103** | 0.410 | 0.405 | 0.350 | 0.375 | **0.244** |
| Dish washer | 1 | 0.127 | 0.135 | 0.100 | 0.114 | 0.075 | 0.453 | 0.495 | 0.293 | 0.329 | 0.051 |
| | 2 | 0.364 | 0.365 | 0.321 | 0.333 | 0.224 | 1.378 | 1.362 | 0.832 | 0.990 | 0.151 |
| | 3 | 0.128 | 0.123 | 0.102 | 0.106 | 0.073 | 1.434 | 1.351 | 0.636 | 0.755 | 0.248 |
| | 4 | 0.089 | 0.086 | 0.105 | 0.095 | 0.087 | 0.065 | 0.049 | 0.163 | 0.123 | 0.025 |
| | **mean** | 0.177 | 0.177 | 0.157 | 0.162 | **0.115** | 0.832 | 0.814 | 0.481 | 0.549 | **0.119** |
| Microwave | 1 | 0.085 | 0.084 | 0.083 | 0.084 | 0.083 | 0.114 | 0.124 | 0.130 | 0.138 | 0.156 |
| | 2 | 0.023 | 0.022 | 0.022 | 0.022 | 0.021 | 0.023 | 0.018 | 0.018 | 0.020 | 0.010 |
| | 3 | 0.028 | 0.028 | 0.013 | 0.014 | 0.013 | 0.468 | 0.453 | 0.037 | 0.055 | 0.023 |
| | 4 | 0.123 | 0.122 | 0.121 | 0.122 | 0.120 | 0.489 | 0.492 | 0.433 | 0.458 | 0.160 |
| | **mean** | 0.065 | 0.064 | 0.060 | 0.060 | **0.059** | 0.273 | 0.272 | 0.154 | 0.168 | **0.087** |

As shown in Table V, using ProfileSR-GAN for upsampling achieves the best performance in most cases. This is because NILM algorithms rely heavily on capturing the load switching signature in the aggregated load profile (e.g., the rising and falling edges, the spikes), which are usually caused by the appliance ON/OFF or cycling activities. Thus, by restoring the high-frequency waveforms, ProfileSR-GAN makes it easier for NILM to identify appliance-level load profiles and energy consumption patterns. The MSE-based SR methods produce over-smoothed HR profiles, which provides fewer waveform signatures for NILM to capture. These results further demonstrate the value of the proposed ProfileSR-GAN model. Figure 15 shows the NILM results for the air conditioner identification as an illustration of the results.

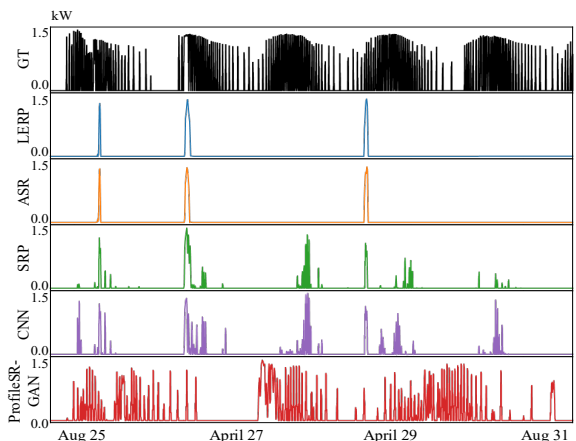

Fig. 15. NILM results comparison for air conditioner based on Seq2Seq algorithm. Data source for the ground truth data set: Pecan Street [11].

## IV. CONCLUSION

In this paper, we propose ProfileSR-GAN, a 2-stage GAN-based method for solving LPSR problems. In the first stage, a GAN-based model is trained to restore the high-frequency components from the low-resolution data. In the second stage, a polishing network is developed to remove unrealistic power fluctuations in the GAN generated high-resolution load profiles. Compared with conventional up-sampling methods, such as interpolation and CNN-based methods, the proposed ProfileSR-GAN achieves superior performance in restoring high-frequency components inside sampling intervals. The overall performance improvements attribute to three aspects: the adversarial training of the GAN-based model, the inclusion of weather data, and the fine-tuning of the polishing network.

The simulation results demonstrate that ProfileSR-GAN achieved 36%-62% improvements in shape-related evaluation metrics compared with the baseline method (i.e., the linear interpolation method). An application of ProfileSR-GAN is presented as a case study to demonstrate that applying ProfileSR-GAN on upsampling can benefit downstream tasks that require the use of high-resolution load profiles. Simulation results show that when using ProfileSR-GAN to upsample the low resolution profiles before conducting NILM, appliance-level activities can be better recognized by the NILM algorithms.

Our future work will be focused on evaluating the performance of ProfileSR-GAN on other time-series data sets such as wind and solar power outputs.

REFERENCES

[1] Y. Wang, Q. Chen, T. Hong, and C. Kang, "Review of smart meter data analytics: Applications, methodologies, and challenges," *IEEE Transactions on Smart Grid,* vol. 10, no. 3, pp. 3125-3148, 2018.

[2] W. Gan *et al.*, "Two-stage planning of network-constrained hybrid energy supply stations for electric and natural gas vehicles," *IEEE Transactions on Smart Grid,* vol. 12, no. 3, pp. 2013-2026, 2020.

[3] K. Hou *et al.*, "A reliability assessment approach for integrated transportation and electrical power systems incorporating electric vehicles," *IEEE Transactions on Smart Grid,* vol. 9, no. 1, pp. 88-100, 2016.

[4] L. Xie, C. Singh, S. K. Mitter, M. A. Dahleh, and S. S. J. J. Oren, "Toward carbon-neutral electricity and mobility: Is the grid infrastructure ready?," *Joule,* vol. 5, no. 8, pp. 1908-1913, 2021.

[5] S. Rahman *et al.*, "Analysis of power grid voltage stability with high penetration of solar PV systems," *IEEE Transactions on Industry Applications,* vol. 57, no. 3, pp. 2245-2257, 2021.

[6] K. Nasrollahi and T. B. Moeslund, "Super-resolution: a comprehensive survey," *Machine Vision and Applications,* vol. 25, no. 6, pp. 1423-1468, 2014, doi: 10.1007/s00138-014-0623-4.

[7] R. Morin, A. Basarab, and D. Kouamé, "Alternating direction method of multipliers framework for super-resolution in ultrasound imaging," in *2012 9th IEEE International Symposium on Biomedical Imaging (ISBI),* 2012: IEEE, pp. 1595-1598.

[8] H. K. Aghajan and T. Kailath, "Sensor array processing techniques for super resolution multi-line-fitting and straight edge detection," *IEEE Transactions on Image Processing,* vol. 2, no. 4, pp. 454-465, 1993.

[9] K. Nguyen, S. Sridharan, S. Denman, and C. Fookes, "Feature-domain super-resolution framework for Gabor-based face and iris recognition," in *2012 IEEE Conference on Computer Vision and Pattern Recognition,* 2012: IEEE, pp. 2642-2649.

[10] V. Kuleshov, S. Z. Enam, and S. Ermon, "Audio super resolution using neural networks," *arXiv preprint arXiv::00853,* 2017.

[11] "Pecan Street Dataport." https://www.pecanstreet.com/dataport/ (accessed 2021).

[12] G. Liu, J. Gu, Z. Zhao, F. Wen, and G. Liang, "Super Resolution Perception for Smart Meter Data," *Information Sciences,* 2020.

[13] X. Xu, D. Sun, J. Pan, Y. Zhang, H. Pfister, and M.-H. Yang, "Learning to super-resolve blurry face and text images," in *Proceedings of the IEEE international conference on computer vision,* 2017, pp. 251-260.

[14] I. Goodfellow *et al.*, "Generative adversarial nets," *Advances in neural information processing systems,* vol. 27, pp. 2672-2680, 2014.

[15] J. Kim, J. K. Lee, and K. M. Lee, "Accurate image super-resolution using very deep convolutional networks," in *Proceedings of the IEEE conference on computer vision and pattern recognition,* 2016, pp. 1646-1654.

[16] Z. Wang, T. Hong, and Buildings, "Generating realistic building electrical load profiles through the Generative Adversarial Network (GAN)," *Energy,* vol. 224, p. 110299, 2020.

[17] Y. Gu, Q. Chen, K. Liu, L. Xie, and C. Kang, "Gan-based model for residential load generation considering typical consumption patterns," in *2019 IEEE Power & Energy Society Innovative Smart Grid Technologies Conference (ISGT),* 2019: IEEE, pp. 1-5.

[18] Y. Zhang, Q. Ai, F. Xiao, R. Hao, T. Lu, and E. Systems, "Typical wind power scenario generation for multiple wind farms using conditional improved Wasserstein generative adversarial network," *International Journal of Electrical Power,* vol. 114, p. 105388, 2020.

[19] Y. Wang, G. Hug, Z. Liu, and N. Zhang, "Modeling load forecast uncertainty using generative adversarial networks," *Electric Power Systems Research,* vol. 189, p. 106732, 2020.

[20] A. Harell, R. Jones, S. Makonin, and I. V. Bajić, "TraceGAN: Synthesizing Appliance Power Signatures Using Generative Adversarial Networks," *IEEE Transactions on Smart Grid,* 2021.

[21] M. Kaselimi, A. Voulodimos, E. Protopapadakis, N. Doulamis, and A. Doulamis, "Energan: A generative adversarial network for energy disaggregation," in *ICASSP 2020-2020 IEEE International Conference on Acoustics, Speech and Signal Processing (ICASSP),* 2020: IEEE, pp. 1578-1582.

[22] C. Ledig *et al.*, "Photo-realistic single image super-resolution using a generative adversarial network," in *Proceedings of the IEEE conference on computer vision and pattern recognition,* 2017, pp. 4681-4690.

[23] P. Cheeseman, B. Kanefsky, R. Kraft, J. Stutz, and R. Hanson, "Super-resolved surface reconstruction from multiple images," in *Maximum Entropy and Bayesian Methods*: Springer, 1996, pp. 293-308.

[24] T. Lukeš *et al.*, "Three-dimensional super-resolution structured illumination microscopy with maximum a posteriori probability image estimation," vol. 22, no. 24, pp. 29805-29817, 2014.

[25] H. Zhang, Y. Zhang, H. Li, and T. S. J. I. T. o. I. p. Huang, "Generative Bayesian image super resolution with natural image prior," vol. 21, no. 9, pp. 4054-4067, 2012.

[26] B. Schölkopf, A. J. Smola, and F. Bach, *Learning with kernels: support vector machines, regularization, optimization, and beyond.* MIT press, 2002.

[27] K. He, X. Zhang, S. Ren, and J. Sun, "Deep residual learning for image recognition," in *Proceedings of the IEEE conference on computer vision and pattern recognition,* 2016, pp. 770-778.

[28] S. Ioffe and C. Szegedy, "Batch normalization: Accelerating deep network training by reducing internal covariate shift," *arXiv preprint arXiv::03167,* 2015.

[29] A. Radford, L. Metz, and S. Chintala, "Unsupervised representation learning with deep convolutional generative adversarial networks," *arXiv preprint arXiv::06434,* 2015.

[30] T. Salimans, I. Goodfellow, W. Zaremba, V. Cheung, A. Radford, and X. Chen, "Improved techniques for training gans," *Advances in neural information processing systems,* vol. 29, pp. 2234-2242, 2016.

[31] M. A. Ranzato, Y.-L. Boureau, and Y. LeCun, "Sparse feature learning for deep belief networks," *Advances in neural information processing systems,* vol. 20, pp. 1185-1192, 2007.

[32] S. R. Bulo, G. Neuhold, and P. Kontschieder, "Loss max-pooling for semantic image segmentation," in *2017 IEEE Conference on Computer Vision and Pattern Recognition (CVPR),* 2017: IEEE, pp. 7082-7091.

[33] U. Ramer and i. processing, "An iterative procedure for the polygonal approximation of plane curves," *Computer graphics,* vol. 1, no. 3, pp. 244-256, 1972.

[34] V. C. Corp. Visual Crossing Weather Services [Online] Available: https://www.visualcrossing.com/weather/weather-data-services#/login

[35] D. P. Kingma and J. J. a. a. p. a. Ba, "Adam: A method for stochastic optimization," 2014.

[36] L. Van der Maaten and G. Hinton, "Visualizing data using t-SNE," *Journal of machine learning research,* vol. 9, no. 11, 2008.

[37] J. Kelly and W. Knottenbelt, "Neural nilm: Deep neural networks applied to energy disaggregation," in *Proceedings of the 2nd ACM international conference on embedded systems for energy-efficient built environments,* 2015, pp. 55-64.

[38] C. Zhang, M. Zhong, Z. Wang, N. Goddard, and C. Sutton, "Sequence-to-point learning with neural networks for non-intrusive load monitoring," in *Proceedings of the AAAI Conference on Artificial Intelligence,* 2018, vol. 32, no. 1.

[39] N. Batra *et al.*, "NILMTK: An open source toolkit for non-intrusive load monitoring," in *Proceedings of the 5th international conference on Future energy systems,* 2014, pp. 265-276.

[40] C. Klemenjak and P. Goldsborough, "Non-intrusive load monitoring: A review and outlook," *arXiv preprint arXiv::01191,* 2016.